\documentstyle{article}
\begin{document}
\hoffset -1.5cm
\def\thefootnote{\fnsymbol{footnote}}

\hfill hep-ph/9507467

\begin{titlepage}
\begin{center}{\Large\bf Systematic Study of Horizontal Gauge Theories}
\vskip.6cm

{\large William A. Ponce$^{1,2}$\footnote{e-mail:
wponce@fisica.udea.edu.co}, Luis A. Wills$^1$, and 
Arnulfo Zepeda$^{3, }$\footnote{On leave 
of absence from Departamento de F\'\i sica, 
Centro de Investigaci\'on y de Estudios Avanzados del IPN, 
Apartado Postal 14-740, 07000 M\'exico D.F., Mexico. e-mail:
zepeda@titan.ific.uv.es}\\
\normalsize 1-Departamento de F\'\i sica, Universidad de Antioquia \\
\normalsize A.A. 1226, Medell\'\i n, Colombia.\\
\normalsize 2-Departamento de F\'\i sica, 
Centro de Investigaci\'on y de Estudios Avanzados del IPN.\\
\normalsize Apartado Postal 14-740, 07000 M\'exico D.F., Mexico.\\
\normalsize 3-Departament de F\'\i sica Te\`orica, Universitat de Val\`encia,
46100 Burjassot, Val\`encia, Spain} 

\end{center}

\vspace{2.cm}

\renewcommand{\baselinestretch}{2}
\setlength{\hsize}{15cm}
\setlength{\vsize}{10in}
 
\hspace{.2cm}

\large
\today

\begin{center}
{\bf ABSTRACT}
\end{center}

\parbox{14cm}%
{We analyze all the possible continuous horizontal gauge groups G$_H$ 
in relation 
with their possibility to explain 
m$_b<<$m$_t$. We assume that the only effective fermionic degrees of 
freedom correspond to the known fermions but allow the possibility of adding 
a right handed neutrino to each family.
We assume that the Higgs fields which generate masses for these fermions,
trough renormalizable Yukawa couplings, transform as an irreducible 
representation of SU(3)$_c\otimes$SU(2)$_L\otimes$U(1)$_Y\otimes$G$_H$.
Under these assumptions we find two U(1)$_H$ or 
U(1)$_{H1}\otimes$U(1)$_{H2}$
models free of anomalies and 
able to guarantee that only the top has a renormalizable mass-generating 
Yukawa coupling.}
\end{titlepage}

\large
\section{Introduction.}
The pattern of fermion masses, their mixing, and the family replication, 
remain as the most outstanding problems of nowadays particle physics. The 
successful Standard Model (SM) based on the gauge group 
SU(3)$_c\otimes$SU(2)$_L\otimes$U(1)$_Y$ can 
tolerate, but not explain the experimental results. Two main features that a 
consistent family theory should provide are\\
(i)-Within each charge sector, the masses increase with family by large 
factors
\begin{center}
$m_u<<m_c<<m_t, \hspace{.5cm} m_d<<m_s<<m_b, \hspace{.5cm} 
m_e<<m_\mu<<m_\tau$,
\end{center}
(ii)-Within each family, the masses are quite 
different. In particular, for the third family we have
\begin{center}
$m_b<<m_t$.
\end{center}
The horizontal survival hypothesis\cite{hsh} was invented in order to 
cope with this hierarchy without putting it by hand in the Yukawa couplings.
According to this hypothesis, 
a certain symmetry should guarantee that at the 
unification scale all the Yukawa terms, with  Yukawa couplings $y_{ff'}$, 
where f and f$^{\prime}$ are 
flavor labels, vanish except for those corresponding to the third family 
for which $y_{tt}\sim y_{bb}\sim  y_{\tau \tau}$.
A different starting point was introduced in Ref. \cite{mhsh}, under 
the name of 
modified survival hypothesis, demanding that all the Yukawa terms 
vanish at the unification scale except the diagonal one of the top quark, 
$y_{tt}\neq 0$.
\vskip.3cm

In this paper we classify the continuous anomaly free horizontal
symmetries that lead to the modified survival hypothesis. We take the number 
of families to be three and  extend
the SM group to SU(3)$_c\otimes$SU(2)$_L\otimes$U(1)$_Y\otimes$G$_H$ and allow
G$_H$ to be any of the subgroups of the most general family symmetry which 
commutes with SU(3)$_c\otimes$SU(2)$_L\otimes$U(1)$_Y$.
We keep the number of ingredients and parameters down to the 
minimum possible assuming that the model does
not contain exotic 
fermions, with the only exception of a possible right-handed neutrino 
state for each family, and assuming that the Higgs fields transform as an 
irreducible representation of 
SU(3)$_c\otimes$SU(2)$_L\otimes$U(1)$_Y\otimes$G$_H$
\vskip.3cm

\noindent
The most general family symmetry which commutes with 
SU(3)$_c\otimes$SU(2)$_L$ is 
\vskip.2cm

 $G =
$U(3)$_q\otimes $U(3)$_u\otimes $U(3)$_d\otimes $U(3)$_l\otimes
$U(3)$_e\otimes $U(3)$_{\nu}$
\vskip.2cm

\noindent
where each factor is defined in the space of 
vectors $\eta = (\eta_1, \; \eta_2 \; \eta_3)$ 
with
$\eta = q, \; u, \; d, \;  l, \; e$, or $\nu$, 

\[
q = \left(\begin{array}{c} u \\d \end{array} \right)_{\alpha L}, \; 
u = u^c_{\alpha L}, \;
d = d^c_{\alpha L}, \;
l = \left(\begin{array}{c} \nu \\ e\end{array} \right)_{L}, \;
e = e^c_{L}, \;
\nu = \nu^c_{L},
\]

\noindent
and where c denotes a charge conjugated field and $\alpha$ is a color
index which will not be displayed  in what follows. In what follows  we 
also omit the helicity index L. Each U(3)$_{\eta} = $SU(3)$_{\eta}\otimes
$U(1)$_{\eta}$ contains a family independent subgroup U(1)$_{\eta}$ and
the SM U(1)$_Y$ factor is contained in
U(1)$_q\otimes $U(1)$_u\otimes $U(1)$_d\otimes $U(1)$_l\otimes
$U(1)$_e\otimes $U(1)$_{\nu}$.
\vskip.3cm

Obviously G is not itself a candidate for a gauged family symmetry since none
of its SU(3)$_{\eta}$ factors, with just one triplet of fermions, 
is anomaly free. In what follows we analyze all
the continuous subgroups G$_H$ of G which are anomaly free and
which allow a (mass generating) Yukawa coupling only for the top quark.
By mass generating Yukawa coupling we mean the coupling to fermions 
of the Higgs 
that develops an SU(2)$_L\otimes$U(1)$_Y$ breaking vacuum expectation 
value (VEV). 
\vskip.3cm

In section 2 we review and extend   
our previous results\cite{po} for the simplest form of G$_H$, namely
G$_H$ = U(1)$_H$. In the following sections we increase the complexity of 
G$_H$. We find two models which satisfy our conditions. They can 
be seen either as U(1)$_H$ models or as  
U(1)$_{H1}\otimes $U(1)$_{H2}$ models. In section 9 we make a brief analysis 
of the phenomenological implications of these two models. In section 10
we summarize our conclusions.

\vskip.3cm

\section{SU(3)$_c\otimes$SU(2)$_L\otimes$U(1)$_Y\otimes$U(1)$_H$  as an 
anomaly-free model.}
A gauge theory is renormalizable if it is free of anomalies\cite{anom}. 
For this particular model only the gravitational\cite{del} and chiral 
anomalies\cite{anom} are present\footnote{The inclusion of gravitational 
constraints has successfully lead to the rederivation of some aspects of 
the Standard Model. See Re. \cite{framoh} and references quoted therein.}.
We 
demand cancellation of these anomalies by the power counting method.
(The alternative of canceling the anomalies by the Green-Schwarz 
mechanism\cite{papa} requires additional assumptions at the string theory
level\cite{ibanezross}.)
\vskip.3cm

There are two different ways of canceling the 
anomalies. One is demanding cancellation of the anomalies 
within each family and the other one is canceling the anomalies among 
families.
\vskip.3cm

\subsection{Cancellation of anomalies within each family.}

Assuming that there are no right-handed neutrinos, using the U(1) charges 
in Table 1a, and demanding freedom from chiral anomalies for 
SU(3)$_c\otimes$SU(2)$_L\otimes$U(1)$_Y\otimes$U(1)$_H$, we get the following 
set of equations ($i = 1, 2, 3$) for the U(1)$_H$ hypercharges:
\vskip.3cm

\begin{eqnarray}
{[{\rm SU}(2)_L]^2U(1)_H }: & & Y_{li}+3Y_{qi}=0, \label{eq:1ch} \\
{[{\rm SU}(3)_c]^2U(1)_H }: & & 2Y_{qi} + Y_{ui}+Y_{di}=0,  \label{eq:2ch}\\
{[U(1)_Y]^2U(1)_H }: & & 2Y_{li}+4Y_{ei}+\frac{2}{3}Y_{qi} + 
\frac{16}{3}Y_{ui} + \frac{4}{3}Y_{di} = 0, \label{eq:3ch} \\
U(1)_Y[U(1)_H]^2: & & -Y^2_{li}+Y^2_{ei}+Y^2_{qi}
-2Y^2_{ui}+Y^2_{di}=0,\label{eq:4ch}\\
{[{\rm grav}]^2U(1)_H }: & & 2Y_{li} + Y_{ei} + 
	3(2Y_{qi} + Y_{ui}+Y_{di})= 0,\label{eq:5ch}\\
{[U(1)_H]^3 }: & & 2Y^3_{li} + Y^3_{ei} + 6Y^3_{qi} 
+ 3Y^3_{ui}+3Y^3_{di} = 0,\label{eq:6ch}
\end{eqnarray}

For a Higgs field with U(1)$_H$ charge Y$_{\phi}$,  a Yukawa coupling for the 
top quark is allowed if 

\begin{equation}
 Y_{q_3}+Y_{u3}=Y_{\phi}  \label{eq:si}
\end{equation}

\noindent
whereas a bottom quark coupling is forbidden if

\begin{equation}
 Y_{q_3}+Y_{d3}\neq -Y_{\phi}=Y_{\phi^\star}. \label{eq:no}
\end{equation}

\noindent
Eqs. ~\ref{eq:si} and ~\ref{eq:no} are however in contradiction with 
Eq.~\ref{eq:2ch}. 
Therefore, if a top quark mass arises at 
tree level, a bottom mass arises as well at the 
same level.

Including right handed neutrinos $\nu^c_{i}$
does not change this conclusion since Eq. 2 stays valid. The only changes 
are in Eqs. 5 and 6 which are now replaced by

\hspace*{-1cm}
\begin{eqnarray} \hspace*{-1cm}
{[{\rm grav}]^2U(1)_H} &:& 2Y_{li}+Y_{ei}+Y_{\nu i} + 
3(2Y_{qi} + Y_{ui}+Y_{di})=0,  \label{eq:5chs}\\
{[U(1)_H]^3} &:&
2Y^3_{li}+Y^3_{ei}+6Y^3_{qi}+3Y^3_{ui}+3Y^3_{di}+Y^3_{\nu i}=0 
\label{eq:6chs}.
\end{eqnarray}

\subsection{Cancellation of anomalies among families.}
If the U(1)$_H$ anomalies are canceled by an interplay among families, 
Eqs. 1 to 6 should be understood with a sum over $i=1,2,3$. Eq. 4 
then reads
\begin{equation}
\sum_i(-Y^2_{li}+Y^2_{ei}+Y^2_{qi}-2Y^2_{ui}+Y^2_{di})=0.
\label{eq:cuad}
\end{equation}

A general class of solutions to the new anomaly cancellation equations
which are linear in $Y_{H}$ is characterized by the
constraints 

\begin{equation}
\sum_{i=1}^3 Y_{\eta i}=0  \label{ys}
\end {equation}

\noindent
for $\eta = q, u, d, l, e$. We will limit ourselves to this type of 
solutions, and within 
this set we will consider only those for which the $u_i$ and $l_i$ U(1)$_H$ 
hypercharges are fixed to satisfy either 

\begin{eqnarray}
Y_{l1}=\delta_1\equiv\delta, \; \; Y_{l2}=\delta_2=-\delta, \; \;
Y_{l3}=\delta_3=0, \nonumber \\
Y_{u1}=\delta_1^\prime\equiv\delta^\prime, \; \; 
Y_{u2}=\delta_2^\prime=-\delta^\prime, \; \; 
Y_{u3}=\delta_3^\prime=0, \label{deltas}
\end{eqnarray}
\noindent
or any set of relations obtained by a permutation 
of the indices $i=1,2,3$. (This guarantees that the ratios of
U(1)$_H$ hypercharges  within the set of fermions of
a given charge are
rational numbers. For solutions with irrational numbers see
Appendix A.) 
These solutions can be divided into four classes 
according to the way the cancellation occurs in
Eq.~\ref{eq:cuad}.
\vskip.3cm

\noindent
\underline{CLASS A} Lepton sector independent of quark sector.\\
$Y_{ei}=Y_{li}=\delta_i$ and $Y_{qi}=Y_{ui}=Y_{di}=\delta_i^\prime,  
i=1,2,3$. A model with a tree-level top quark mass arises if 
$Y_{\phi}=Y_{qi}+Y_{uj}$ for some $i$ and $j$. There are five 
different models in this class characterized by 
Y$_{\phi}=\pm 2\delta^\prime,\pm\delta^\prime$ and 0 respectively. 
Simultaneously 
a tree-level bottom
mass arises if there exist family indices $k$ and $l$ for which 
$Y_{qk}+Y_{dl}=-Y_{\phi}$. 
For example, if $Y_{\phi}=2\delta^\prime$ then $i=j=1$ and $k=l=2$ satisfy 
both conditions; this is signaled in Table 2 by the entries 
$(1,1)_u$ and $(2,2)_d$ in the Class A column and the $2\delta^\prime$ row. 
The fact that in Table 2 there is at least one $d$-type entry for every 
$u$-type one for all the five models of Class A means that none of them is 
viable. 
(This stems from the equality  $Y_{di}=Y_{ui}$
in Class A). 
\vskip.3cm

\noindent
\underline{CLASS B} Doublets independent of singles.\\
$Y_{qi}=Y_{li}=\delta_i$ and $Y_{ui}=Y_{di}=Y_{ei}=\delta_i^\prime, 
i=1,2,3$. There are nine different models in this class characterized by 
Y$_{\phi}=\delta\pm\delta^\prime, \; -(\delta\pm\delta^\prime), \;  
\pm\delta, \; 
\pm\delta^\prime$, and 0 respectively. Again  none of 
these models is viable ($Y_{di}=Y_{ui}$ in Class B also).
\vskip.3cm

\noindent
\underline{CLASS C} Interplay between doublet leptons and singlet quarks.\\
$Y_{di}=Y_{li}=\delta_i$ and $Y_{qi}=Y_{ui}=Y_{ei}=\delta_i^\prime, 
\; i=1,2,3$. There are now eleven different models in this class characterized 
by 
Y$_{\phi}=\pm 2\delta^\prime, \; \delta\pm\delta^\prime, \; 
-(\delta\pm\delta^\prime), \;  \pm\delta, \; \pm\delta^\prime$ and 0. 
As can be seen 
from Table 2, for 
$\delta\neq 0,\pm\delta^\prime,\pm 2\delta^\prime, \pm 3\delta^\prime$ and 
$\delta^\prime\neq 0$ there are two models in which only one $u$-type mass and 
none $d$-type one develops at tree-level. These models are:\\
\underline{MI}. Where the Higgs field has
(U(1)$_Y$,U(1)$_H$) hypercharges equal to $(-1,2\delta^\prime)$.\\
\underline{MII}. Where the Higgs field has
(U(1)$_Y$,U(1)$_H$) hypercharges equal to $(-1,-2\delta^\prime)$.\\
The rest of the models in this class are non-viable because a 
bottom quark mass arises at tree level. The U(1)$_H$ quantum numbers of the 
Yukawa terms in the models MI and MII are displayed in Appendix B.
\vskip.3cm

\noindent
\underline{CLASS D} Same U(1)$_H$ hypercharge for the whole family.\\
 This is a particular case of 
classes A,B, and C above, for which $\delta_i=\delta_i^\prime$, which in turn 
implies\cite{koca}
$Y_{qi}=Y_{ui}=Y_{di}=Y_{li}=Y_{ei}=\delta_i $.
As far as the quark mass spectrum is concerned this class is equivalent to 
class A. 
\vskip.3cm

We may include 
right-handed neutrino fields within the above scheme either 
by setting $Y_{\nu 1}=-Y_{\nu 2}=\delta , \; 
Y_{\nu 3}=0$ (or permutations of the 
indices 1,2,3), or by imposing $Y_{\nu i}=0$ (which is one of
the ingredients for the see-saw mechanism\cite{seesaw}).

\section{SU(3)$_c\otimes$SU(2)$_L\otimes$U(1)$_Y\otimes$U(1)$_{H1}\otimes$
U(1)$_{H2}$.}

Let us consider first the case where U(1)$_{H1} = $U(1)$^{L}$ and
U(1)$_{H2} = $U(1)$^{R}$.
There are again two different ways of canceling the anomalies, the 
cancellation  within each family, and the cancellation among families.

\subsection{Cancellation of anomalies within each family.}
Without right-handed neutrinos, the 
only anomaly-free solution corresponds to the trivial one 
Y$^L$=Y$^R=0$. Let us see why. 
\vskip.3cm

The [SU(3)$_c]^2$U(1)$_{HL}$ constraint is Y$_{qi}^L=0$. The 
[SU(2)$_L]^2$U(1)$^L$ constraint is Y$_{li}^L$ + 3Y$^L_{qi}=0$, 
which combined with the previous result implies Y$_{li}^L=0$. Hence, 
Y$^L=0$ as stated. (This result still holds when we include 
right-handed neutrinos which are singlets under SU(2)$_L$ and SU(3)$_c$). 
Now the [SU(3)$_c]^2$U(1)$^R$ constraint implies Y$_{ui}^R$ + Y$_{di}^R=0$. 
The [grav]$^2$U(1)$^R$ constraint implies 
Y$_{ei}^R$ + 3Y$^R_{ui}$ + 3Y$^R_{di}=0$, which combined with the 
[SU(3)$_c]^2$U(1)$^R$ constraint 
gives Y$_{ei}^R=0$. Now if we combine the last result with the 
[U(1)$^R]^2$U(1)$_{Y}$ constraint $(Y^R_{ei})^2-2(Y^R_{ui})^2 + 
(Y^R_{di})^2=0$, we get 
Y$_{di}^R$=Y$_{ui}$=Y$_{ei}^R$=0 as anticipated 
above.
\vskip.3cm

When we include the right-handed neutrinos we still have Y$^L=0$ as 
commented in the previous paragraph. The anomaly cancellation 
equations for Y$^R$ are then given by equations 
~\ref{eq:1ch} - ~\ref{eq:4ch} and ~\ref{eq:5chs} and ~\ref{eq:6chs} 
with $Y_{li}=Y_{qi}=0$. The solution to those equations is 
$Y^R_{ui}=Y^R_{\nu i}=-Y^R_{di}=-Y^R_{ei}=\kappa_i$, where $\kappa_i$ 
are three arbitrary numbers. 
Therefore a Higgs field with 
$Y^R_{\phi}=\kappa_3$ will produce Dirac masses for the top and bottom 
quarks, and for the $\tau$ and $\nu_\tau$ leptons as well.

\subsection{Cancellation of anomalies among families.}
Using the U(1)$^L$ and U(1)$^R$ charges in Table 1a we get the 
following set of equations to be satisfied simultaneously:
\vskip.3cm

\noindent
$\begin{array}{lll}
[{\rm SU}(2)_L]^2 U(1)^{L(R)}: & \sum_i (Y^L_{li}+3Y^L_{qi})=0 
&  \\
{[{\rm SU}(3)_c]^2U(1)^{L(R)}}: & \sum_i Y^L_{qi}=0 &
\sum_i (Y^R_{ui}+Y^R_{di})=0\\
{[U(1)_Y]^2U(1)^{L(R)}}: & \sum_i (3Y^L_{li}+Y^L_{qi})=0 &
\sum_i (3Y^R_{ei}+4Y^R_{ui} + Y^R_{di}) = 0\\
U(1)_Y[U(1)^{L(R)}]^2: & \sum_i [(Y^L_{qi})^2 -(Y^L_{li})^2]=0 &
\sum_i [(Y^R_{ei})^2-2(Y^R_{ui})^2+(Y^R_{di})^2]=0 \\
{[{\rm grav}]^2U(1)^{L(R)}}: &  \sum_i (3Y^L_{qi}+Y^L_{li})=0 &
\sum_i (Y^R_{ei}+3Y^R_{ui}+3Y^R_{di}) = 0  \\
{[{\rm U}(1)^{L(R)}]^3 }: & \sum_i [3(Y^L_{qi})^3+(Y^L_{li})^3]=0 &
\sum_i [(Y^R_{ei})^3 + 3(Y^R_{ui})^3+3(Y^R_{di})^3] = 0.
\end{array}$
\vskip.3cm

The mixed anomalies related to 
U(1)$^L$[U(1)$^{R}]^2$ and U(1)$^{R}$[U(1)$^L]^2$
trivially vanish due to the fact that for every SM multiplet one of the two 
hypercharges $Y^L$ or $Y^R$ is always zero (see Table 1a).
\vskip.3cm

\noindent
Contrary to the case with a single U(1)$_H$ factor group, the constraint given
by eq. (\ref{ys}), follows fron the above anomaly cancellation eqs. Furthermore,
the quadratic anomaly cancellation eqs. demand now that
the case of cancellation of anomalies among doublets be independent
of the cancellation among singlets. Therefore we set

\centerline{$Y^L_{l1}=\delta_1\equiv\delta, \; \; Y^L_{l2}=\delta_2=-\delta, 
\; \;
Y^L_{l3}=\delta_3=0, \; \; Y^L_{qi}=Y^L_{li},$}
\vskip.3cm

\centerline{$Y^R_{u1}=\delta_1^\prime\equiv\delta^\prime, \; \;
Y^R_{u2}=\delta_2^\prime=-\delta^\prime, \; \; 
Y^R_{u3}=\delta_3^\prime=0,\;  \;
Y^R_{ei}=Y^R_{di}=Y^R_{ui}=\delta_i^\prime,$}
\vskip.3cm

\noindent
or any set of relations obtained from them by a permutation of the indices 
$i=1,2,3$ (for other solutions see Appendix A). 
These solutions are similar to the solutions in Class B in 
section 2.2, hence the same conclusions follow. 

\subsection{U(1)$_{H1} =$U(1)$^{quarks}$, U(1)$_{H2}$ = U(1)$^{leptons}$.}
Again, if we demand cancellation of anomalies within each family, the linear
constraints imply Y$^{quarks} = 0$ for all fermions. When anomalies are
canceled among families the situation is similar to that of class A models 
in the case G$_H$ = U(1)$_H$. Setting 

$$
 Y^{quarks}_{qi}=Y^{quarks}_{ui}=Y^{quarks}_{di}=\delta^{\prime}_i
$$

\noindent
and 
 
 $$
Y^{leptons}_{li}=Y^{leptons}_{ei}=\delta_i
$$

\noindent
we are led again to the conclusion that there are no viable models in 
this case.

\subsection{U(1)$_{H1} =$U(1)$^{que}$, U(1)$_{H2}  =$U(1)$^{dl}$.}
Again, nontrivial horizontal hypercharges are compatible with anomaly 
cancellations only when these are realized among families.
\vskip.3cm

Setting 
 
$$
 Y^{que}_{qi}=Y^{que}_{ui}=Y^{que}_{ei}=\delta^{\prime}_i,
$$

 $$
Y^{dl}_{li}=Y^{dl}_{di}=\delta_i,
$$

\noindent
we obtain the models MI and MII of section 2 with Y$^{que}_{\phi}
= \pm 2\delta^{\prime}, \; \; Y^{dl}_{\phi}=0$.
For the Yukawa terms to be invariant under the full G$_H =$
U(1)$^{que}\otimes$U(1)$^{dl}$ a vanishing entry in the matrices of quantum 
numbers displayed in Appendix B should be obtained without the interplay
of $\delta$ with $\delta^{\prime}$. Therefore the only condition that on
$\delta^{\prime}$ imposes the requirement that an invariant Yukawa term 
be allowed only for the top quark is just $\delta^{\prime} \neq 0$.
\vskip.3cm

We close this section by remarking that a  glance to Eq.~\ref{eq:cuad}
should convince the reader that it is not viable to consider additional
U(1) factors in G$_H$, G$_H$ = 
U(1)$_{H1}\otimes$ U(1)$_{H2}\otimes$ U(1)$_{H3}\otimes$ ...That is, 
there is no other way of canceling the U(1)$_Y$[U(1)$_{H1}$]$^2$ anomalies.

\section{SU(3)$_c\otimes$SU(2)$_L\otimes$U(1)$_Y\otimes$SU(2)$_H$.}
All the possible models for this group 
have been exhaustively analyzed in Ref.\cite{volkas}. We use here some of 
their results. The anomaly free SU(2)$_H$ spectra in the absence 
of right-handed neutrinos shows that from the $9\times 27$ possible 
arrangements of representations in the gauge group only 14 satisfy the 
chiral and global\cite{witten} anomaly constraints 
(there is not a gravitational anomaly 
for SU(2)$_H$, and the global anomaly vanishes if the theory contains 
an even number of SU(2)$_H$ doublets\cite{witten}). 
These fourteen possible models are presented in Eq.3 
of Ref.\cite{volkas} and 
analyzed in Sections. II, III, and in the summary of the tree level results 
presented in Table IV of the same reference. From that Table (which is correct 
up to minor details) we see that none of the possible models is able to 
produce a tree level rank one mass matrix for the up quark sector 
simultaneously with a rank zero mass matrix for the down sector. In 
more detail the situation is the following:
\vskip.3cm

From Eq. 3 in Ref.\cite{volkas} we see that there are only 11 different 
quark arrangements to be considered in the 14 theories enumerated 
(Q$_1$, Q$_8$ and Q$_{21}$ appear twice) 
where the possible quark SU(2)$_H$ representations 
Q$_i, i=1-27$ are in Table 4a  considering that in the 
table all the representations belong to the same horizontal group SU(2)$_H$. 
Now, SU(2)$_H$ does not act on the quark fields 
in Q$_{21}$, hence it has for the quark sector the same information 
than the SM has and we ignore it. We also ignore
Q$_{13}$ since in this arrangement  SU(2)$_H$ does not act on the 
up quark sector. 
Then we should analyze only the following SU(2)$_H$ 
structures: Q$_1$, Q$_4$, Q$_6$, Q$_8$, Q$_{10}$, Q$_{11}$, Q$_{12}$, 
Q$_{16}$, and Q$_{18}$. 
\vskip.3cm

Since a mass term for the up quark sector is of the form 
$\langle\phi^{\dagger} \rangle u^c q $, and that for the down 
quark sector is of the form 
$\langle\phi^{\prime \dagger} \rangle d^c q$,
 and since 
SU(2)$_{H}$ is a (pseudo-) real  group, it is obvious that when 
$u^c$ and  $d^c$ are in the same representation of SU(2)$_H$, a Higgs field 
$\phi$ which produces a mass matrix for the up sector will produce, via 
$\phi^c=-\sigma_{2L}\phi^\star$ or
$\phi^c=-\sigma_{2L}\sigma_{2H}\phi^\star$, the same mass matrix for the 
down sector (up to a Yukawa coupling constant). Therefore
the structures Q$_1$, Q$_4$, Q$_8$, Q$_{11}$, Q$_{12}$ and Q$_{18}$ are not 
viable. 
We are then left only with the structures Q$_6$, Q$_{10}$ and Q$_{16}$ 
to be analyzed.
\vskip.3cm

Q$_{16}$ was analyzed in full detail in Ref.\cite{volkas}. 
The SU(3)$_c\otimes$SU(2)$_H\otimes$
SU(2)$_L\otimes$U(1)$_Y$ quantum 
numbers (QNs) for the quark sector in this model are
$q\sim (3,2,2)_{1/3}\oplus(3,1,2)_{1/3}$,
$u^c\sim (\overline{3},2,1)_{-4/3}\oplus(\overline{3},1,1)_{-4/3}$, 
$d^c\sim 3(\overline{3},1,1)_{2/3}$.
An up quark mass term in Q$_{16}$ has SU(2)$_H$ QNs according to the 
product $(2+1)\times(2+1)=3+2(twice)+1(twice)$. Since a 
down quark mass term has QNs $(2+1)\times 1=2+1$, a Higgs field 
belonging to the representations 1 or 2 of SU(2)$_H$ 
will produce non zero mass matrices for the up and down sectors simultaneously. 
A Higgs field  $\phi_{k}$ belonging to the representation 3 of SU(2)$_H$ 
generates a mass matrix only for the up quark sector
of the form $\langle \phi_{k}^{\dagger}\rangle \sum^2_{c,b=1} u^c_a 
(\sigma^k)_{ab}q_b $, where each $\phi_{k}$ is an SU(2)$_L$ doublet
with VEV of the form 
$(x_{k}, 0)^T $.
With all\footnote{At the scale $\langle \phi_k \rangle$ SU(2)$_H$ is 
aready broken and we cannot use it to orientate $\langle \phi_k \rangle$
in the H space}
the $x_{k} \neq 0$ the rank of the up quark mass matrix is two, 
with 
$m_t=m_c$, which is unphysical\cite{volkas}. Hence
Q$_{16}$ is not able to explain the known quark mass spectrum.
\vskip.3cm

Now for Q$_6$, the 
SU(3)$_c\otimes$SU(2)$_H\otimes$SU(2)$_L\otimes$U(1)$_Y$ QNs for the quark 
sector are
$q\sim (3,2,2)_{1/3}\oplus(3,1,2)_{1/3}$,
$u^c\sim (\overline{3},2,1)_{-4/3}\oplus(\overline{3},1,1)_{-4/3}$,
$d^c\sim (\overline{3},3,1)_{2/3}$.
An up quark mass term has SU(2)$_H$ QNs 3, 2 or 1 as for Q$_{16}$, and a 
down quark mass 
term has SU(2)$_H$ QNs $(2+1)\times 3=4+2+3$. Hence, a Higgs field 
belonging to the representation 2 or 3 of SU(2)$_H$ 
produces masses both for the up and down quark sectors. A singlet 
Higgs field under SU(2)$_H$ generates a mass matrix only for the up sector, 
but it couples to the three families producing a rank three mass matrix 
and generating $m_t,m_c$ and $m_u$ at tree-level. Therefore 
Q$_6$ is discarded. 
\vskip.3cm

Finally, the SU(3)$_c\otimes$SU(2)$_H\otimes$SU(2)$_L\otimes$U(1)$_Y$ 
QNs for Q$_{10}$ are
$q\sim (3,3,2)_{1/3}$,
$u^c\sim (\overline{3},3,1)_{-4/3}$,
$d^c\sim 3(\overline{3},1,1)_{2/3}$.
Now an up quark mass term has SU(2)$_H$ QNs $3\times 3=5+3+1$ and a down quark 
mass term is a triplet under SU(2)$_H$. Thus a Higgs field belonging to the 
representation 3 of SU(2)$_H$ will 
generate mass matrices for the up and down sectors at the same time, but a 
Higgs field in the representation 1 or 5 of SU(2)$_H$ 
will produce tree-level masses only for 
the up sector. However the mass matrices generated in both cases are rank 
three\cite{volkas}, producing three-level masses for the three families. 
Therefore this possibility, the last one in the case without right handed
 neutrinos, is also discarded.
\vskip.3cm

When the right-handed neutrinos are included in the spectrum (one for each 
family), the number of models satisfying the anomaly constraints become 35 
(see equation (35) in Ref.\cite{volkas}). These 35 models correspond to the 14 
structures of the neutrinoless case with the three neutrinos accommodated in the 
representation 3 of SU(2)$_H$, plus the same 14 structures of the neutrinoless 
case with the three neutrinos accommodated in the representation 
$1\oplus 1\oplus 1$ of SU(2)$_H$, plus seven more structures with the neutrinos 
accommodated in the representation (2+1) of SU(2)$_H$, and the quark 
spectrum given by (Q$_1$, Q$_3$, Q$_5$, Q$_8$, Q$_{17}$, Q$_{19}$, and 
Q$_{21}$). The structures Q$_1$, Q$_8$, and  Q$_{21}$ where analyzed 
and excluded in the previous paragraphs. Also Q$_5$ and Q$_{17}$ are such 
that $u^c$ 
and $d^c$ are in the same SU(2)$_H$ representation, and in Q$_{19}$ 
SU(2)$_H$ does not act on the up quark sector at all. 
We are thus left only with Q$_3$ to be analyzed here. 
\vskip.3cm

The SU(3)$_c\otimes$SU(2)$_H\otimes$SU(2)$_L\otimes$U(1)$_Y$ 
QNs for the quark sector in Q$_{3}$ are
$q\sim (3,3,2)_{1/3}$,
$u^c\sim (\overline{3},3,1)_{-4/3}$,
$d^c\sim (\overline{3},2,1)_{2/3}\oplus (\overline{3},1,1)_{2/3}$.
An up quark mass term has SU(2)$_H$ QNs 5,3,1 as in  Q$_{10}$, and a down quark 
mass term is of the form $3\times (2+1)\sim 4+2+3$. A Higgs field belonging 
to representation 3 of 
SU(2)$_H$ generates thus masses for the up and down quark sectors 
simultaneously. A Higgs field in a 1 or 5 representation of SU(2)$_H$ 
generates a mass matrix only for the up quark sector, but of 
rank 3 \cite{volkas}. Therefore Q$_3$ is also ruled out.
\vskip.3cm

The conclusion is that 
SU(3)$_c\otimes$SU(2)$_H\otimes$SU(2)$_L\otimes$U(1)$_Y$ can not explain 
why $m_t>>m_b$.

\section{SU(3)$_c\otimes$SU(2)$_L\otimes$U(1)$_Y\otimes$
SU(2)$_{HL}\otimes$SU(2)$_{HR}$.}
Gauge anomaly 
cancellation is satisfied if the [SU(2)$_{HL}]^2$U(1)$_Y$ and 
[SU(2)$_{HR}]^2$U(1)$_Y$ anomalies vanish 
while global anomaly cancellation requires that there be an 
even number of SU(2)$_{HL}$ and of SU(2)$_{HR}$ doublets. 
Tables 3 and 4a show the values of the anomalies for the various 
possible ways of assigning representations to the particle types.
\vskip.3cm

Comparing the values from these tables, and without including right-handed 
neutrinos, only six theories satisfying gauge and global anomaly 
cancellation appear. They are: (L$_1$+Q$_1$, L$_3$+Q$_4$, L$_5$+Q$_{12}$, 
L$_6$+Q$_{11}$, L$_7$+Q$_{18}$, L$_9$+Q$_{21}$). The last of these models is 
just the SM and therefore SU(2)$_{HL}\otimes$SU(2)$_{HR}$ does not act upon it. 
Notice also from Table 4a that for the other five quark structures 
Q$_1$, Q$_4$, Q$_{11}$, Q$_{12}$ and Q$_{18}$ (and also for Q$_{21}$ if we 
wish) the up and down sectors belong to the same SU(2)$_{HR}$ representation, 
and then, a Higgs field which produces a non-zero mass matrix for the up 
sector will produce a non-zero mass matrix for the down sector as well. 
Therefore, 
all of them are ruled out.
\vskip.3cm

When we include the three right-handed neutrinos we see that they do not 
contribute to the gauge anomaly (Y$_{SM}=0$ for $\nu^c_{i}$), and their 
contribution to the global anomaly will also be zero if the neutrinos 
transform either as three singlets or as a triplet under SU(2)$_{HR}$ (N$_1$ 
and N$_3$ respectively in what follows). Now, when they 
transform as a singlet plus a doublet (N$_2$), they contribute with an extra 
doublet to the global anomaly. There are then a total 
of 14 models without anomalies, the six structures of the neutrinoless 
case each one added with 
N$_1$ or with N$_3$, plus the following two new structures: 
(L$_4$ + Q$_8$ + N$_2$) and (L$_8$ + Q$_{17}$ + N$_2$). The first 12 models 
were already ruled out.
\vskip.3cm

The two new quark structures Q$_8$ and Q$_{17}$ are also 
such that the up and down quark sectors belong to the same representation of 
SU(2)$_{HR}$ (2+1). Therefore they are also ruled out. Our conclusion is 
that SU(3)$_c\otimes$SU(2)$_L\otimes$U(1)$_Y\otimes$
SU(2)$_{HL}\otimes$SU(2)$_{HR}$ is not able to explain why $m_t>>m_b$.
\vskip.3cm

\section{SU(3)$_c\otimes$SU(2)$_L\otimes$U(1)$_Y\otimes$
SU(2)$_{H1}\otimes$SU(2)$_{H2}$.}

In section 3.4 it was shown that U(1)$^{que} \otimes 
$U(1)$^{dl}$ satisfies our requirements. We may then ask whether a different
assignment of the $\eta$ fields to SU(2)$_{H1}$ and SU(2)$_{H2}$, other than 
separating left-handed fields from right-handed ones, would lead to a viable
model.
\vskip.3cm

We start by looking for the representations content of the $q$, $u$ and 
Higgs fields that give rise to a rank-one up-quark mass matrix. Let then 
$n_{\eta} = (1+1+1)$, (2+1), or 3 be the representation content of $\eta =
q$ or $u$ under some fixed SU(2)$_{H1}$ group. There are nine possibilities
for $n = (n_q, \, n_u)$. Although the rank of the up quark mass matrices 
that are allowed in each case has been already discussed in the previous 
section, we follow here a slightly different path in order to be able to leave
unspecified the SU(2)$_{H1}$ group. It is straightforward to compute
the rank of the up-quark mass matrices in each case but  we can also 
read it from the tables IV and V of Ref. \cite{volkas}. The results are listed 
in table 4b together with a Q$_i$ arrangement which contains the same 
$(n_q, \, n_u)$ entry and with the Higgs representations that leads
to a non zero
rank.
\vskip.3cm

After we discard from table 4b the cases with rank $> 1$, we are left with 8
($n_q$, $n_u$, $n_{\phi}$) =  $n_A$  arrangements. Given 
$n_q$ and $n_{\phi}$ a down-quark mass term is forbidden if 
$n_{\phi}$ is not in the set of irreducible representations 
in the complete reduction of 
$n_q \times n_{d}$. Table 4c lists the above mentioned $n_A$
together with the values of $n_{d}$ which forbid a down-quark mass term and 
with the 
corresponding [SU(2)$_{H1}$]$^2$U(1)$_Y$ anomalies. As we can see, all the cases
are anomalous and we have to include some leptons in SU(2)$_{H1}$. The 
contribution to the anomaly from $n_l = (1+1+1+)$, 2+1, 3 is, however,
= 0, -2, -8 while that of $n_e =(1+1+1+)$, 2+1, 3 is 0, 2, 8 and that of 
$n_{\nu} = 0$. Therefore it is impossible to cancel the anomalies listed 
in table 4c.
and none of the  SU(2)$_{H1}\otimes $SU(2)$_{H2}$ models is viable. 
The same reasoning applies to 
SU(2)$_{H1}\otimes $U(1)$_{H1}\otimes $SU(2)$_{H2}\otimes $U(2)$_{H2}$ or 
any other
G$_H$ such that SU(2)$_{H1}\otimes $SU(2)$_{H2} \subset $ G$_H \subset $ G. 
For the sake of 
illustration and completeness we discuss in the following two sections
two cases with other higher symmetry G$_H$ groups.
\vskip.3cm

\section{SU(3)$_c\otimes$SU(2)$_L\otimes$U(1)$_Y\otimes$
SU(3)$_{H}$.}
This model with quarks and leptons in the vectorlike $3+\bar{3}$ 
representation of SU(3)$_H$ 
was introduced for the first time in the literature in 
Ref.\cite{yana} and analyzed later in Refs.\cite{zupa}. In Ref.\cite{ray} 
the known quarks and leptons where assigned to the chiral 3+3 representation 
of SU(3)$_H$ canceling the anomalies with mirror fermions, and recently in 
Ref.\cite{bere}, the anomalies in this last representation were canceled 
in a more general way.
\vskip.3cm

[SU(3)$_c$, SU(2)$_L$] multiplets may belong to the
 1, 3 or $\bar{3}$ representation of SU(3)$_H$. On the other hand, 
since SU(3) is not a real group, the cancellation 
of the [SU(3)$_{H}]^3$ and 
[SU(3)$_H]^2$U(1)$_Y$ anomalies is nontrivial; in particular
the cancellation of the first 
anomaly is achieved only when the number of SU(3)$_H$ triplets equals the 
number of antitriplets. Finally, since there are 18 different quark fields 
(36 Weyl states) but only 6 different lepton fields (12 or 9 Weyl states 
depending on whether we include or not the right-handed neutrino states), 
then it is not always possible to cancel the quark anomalies with the 
lepton anomalies. 
\vskip.3cm

In Table 5 we include all the possible SU(3)$_H$ fermion field 
representation assignments which are free of the SU(3)$_H$ gauge anomalies, 
together with their [SU(3)$_H]^2$U(1)$_Y$ anomaly value. From 
this table we find that only the models M$_{3}$, M$_{4}$, M$_{8}$, M$_{13}$, 
M$_{14}$, M$_{17}$, M$_{18}$, and  M$_{19}$,  are safe. (Notice that 
without right-handed neutrino states, the only anomaly free model is M$_{4}$ 
and that for M$_{13}$ and M$_{14}$ SU(3)$_H$ does not act on the left-handed 
fields at all).
M$_{19}$, M$_{3}$ and M$_{8}$ are not adequate candidates 
since
M$_{19}$ is just the SM,
SU(3)$_H$ does not act on the quark sector of M$_{3}$, and
SU(3)$_H$ does not act on the up quark sector of M$_{8}$.
\vskip.3cm

In M$_{4}$, a mass term for the up sector has SU(3)$_H$ QNs 
$3\times\bar{3}=8+1$, and a mass term for the down sector has the same 
SU(3)$_H$ QNs. Since both representations 8 and 1 are real, a Higgs field 
$\phi$ which produces a mass matrix for the up quark sector will also produce
a down quark mass matrix via $i\sigma_{2L}\phi^\star$. This argument is also 
valid for M$_{17}$ and M$_{18}$ (see Refs.\cite{yana,zupa}). Therefore 
M$_{4}$, M$_{17}$ and  M$_{18}$ can not explain why $m_b<<m_t$.
\vskip.3cm

Now, for M$_{13}$ and M$_{14}$ 
a Higgs field $\phi^\prime$ able to produce a mass term for 
the up sector $\langle\phi^\prime\rangle^\dagger u^c q$ must be in 
the representation $\bar{3}$ of SU(3)$_H$. But such a Higgs field automatically 
produces a down quark mass term 
$\langle\stackrel{\sim}{\phi}^\prime\rangle^\dagger d^c q$ via 
$\stackrel{\sim}{\phi}^\prime\equiv i\sigma_{2L}\phi^{\prime\star}$. 
Thus again we conclude that SU(3)$_H$ by 
itself can not explain why $m_b<<m_t$.

\section{SU(3)$_c\otimes$SU(2)$_L\otimes$U(1)$_Y\otimes$
SU(3)$_{HL}\otimes$SU(3)$_{HR}$.}

Since SU(3)$_{HL}$ would act only on $q_{i}$ and 
$l_{i}$, the SU(3)$_{HL}$ 
safe representations are only those for which the anomalies in the 
quark sector cancel exactly the anomalies in the lepton sector, which is 
impossible when $q$ and/or $l$ are in a nontrivial representation 
(3 or $\overline{3}$)
due to the fact that in this case there would be six 
SU(3)$_L$ representations in q 
(due to color) while only two in $l$.
Therefore SU(3)$_{HL}\otimes$SU(3)$_{HR}$ is equivalent to a single SU(3)$_{H}$ 
which accommodates the left-handed fields $q_{i}$ and $l_{i}$ in 
the 
representation 1+1+1. This information can be extracted from Table 5 
identifying SU(3)$_{H}$ as SU(3)$_{HR}$. 
\vskip.3cm

From Table 5 we read that when $q_i$ and $l_{i}$ are in the representation 
1+1+1 of SU(3)$_{HL}$ only
M$_{13}$ and M$_{14}$ are 
anomaly free.
But those two models are also ruled out 
by the same reason that they were ruled out in the previous 
section. Hence SU(3)$_c\otimes$SU(2)$_L\otimes$U(1)$_Y\otimes$
SU(3)$_{HL}\otimes$SU(3)$_{HR}$ by itself can not explain why $m_b<<m_t$.

\section{Brief analysis of MI and MII models.}
In the previous sections we have analyzed all the possible horizontal  
gauge models able to accommodate three families. From them we selected 
those models able to generate a rank one tree-level mass matrix for the 
up quark sector and a rank zero tree-level mass matrix for the down quark 
sector. We have found only two candidates named MI and MII, both 
of them with the horizontal structure U(1)$_H$ or U(1)$^{que} \otimes 
$U(1)$^{dl}$. 
\vskip.3cm

Table Ib depicts explicitly the U(1)$_H$ 
(or U(1)$^{que} \otimes $U(1)$^{dl}$) 
hypercharges of fermions in 
these models.
These hypercharges are then responsible for the quantum numbers of all 
possible (mass generating) Yukawa terms displayed in Appendix B. 
As can be seen, only one Yukawa
term has zero horizontal hypercharge, 

\[ {\rm MI:} \; \; \; 
\phi^\dagger u^c_{1}\left( \begin{array}{c} u\\d \end{array} \right)_{1} \]

\[ {\rm MII:} \; \; \; 
\phi^\dagger u^c_{2}\left( \begin{array}{c} u\\d \end{array} \right)_{2}. \]

In what follows we briefly comment on some aspects of these models.

\subsection{The symmetry breaking chain.}
For simplicity, let us introduce two Higgs fields $\phi_H$ and $\phi_{SM}$, 
both of them developing vacuum expectation values.
$\langle\phi_H\rangle\sim M_H$ and 
$\langle\phi_{SM}\rangle\sim  M_Z$, such that

\[ SU(3)_c\otimes SU(2)_L\otimes U(1)_Y\otimes G_H 
\stackrel{M_H}{\longrightarrow} SU(3)_c\otimes SU(2)_L\otimes U(1)_Y \]

\[\stackrel{M_Z}{\longrightarrow} SU(3)_c\otimes U(1)_Q, \]

\noindent
where $G_H$ = U(1)$_H$ or U(1)$^{que} \otimes 
$U(1)$^{dl}$ and where  $Q=T_{3L}+Y/2$. 

\vskip.2cm

The SU(3)$_c\otimes$SU(2)$_L\otimes$U(1)$_Y\otimes$U(1)$_H$ QNs for 
$\phi_H$ and $\phi_{SM}$ are $(1,1,0,\pm (3\delta^\prime + \delta)/2)$ and 
$(1,2,-1,\pm 2\delta^\prime)$ respectively, where the U(1)$_H$ hypercharges  
are chosen for further purposes, and the upper and down signs 
are related to MI and MII respectively. 
\vskip.3cm

The required Higgs system in this section is minimal in the sense that there 
is a Higgs SU(2)$_L$ doublet associated with the SM mass scale, and a singlet 
which provides the desired mass of the U(1)$_H$ gauge boson Z$^\prime$, heavy 
enough to avoid conflict with experiments.

\subsection{Bottom quark mass.}
To generate a bottom quark mass matrix different from zero in MI 
or MII further 
ingredients must be added to the models. For this purposes we introduce 
two new Higgs fields\cite{ma} 
$\phi^{(1)}$ and $\phi^{(2)}$ which do not develop VEVs and with QNs 
given by $\phi^{(1)}(3,1,-2/3,\mp 2\delta^\prime)$ and 
$\phi^{(2)}(3^{\star},1,2/3,\mp(\delta+\delta^\prime))$. With these new 
Higgs fields other 
Yukawa terms are allowed in the Lagrangian,

\[{\cal L}^\prime=\epsilon_{\alpha \beta \gamma}(h_1 q_{1}^{\alpha }
\sigma_{2L}
q_{1}^\beta \phi^{(1) \gamma}+h_2u_{1}^{c \alpha} 
d_{1}^{c \beta} \phi^{(2) \gamma}) + h.c.,\]

\noindent
where $\sigma_{2L}$ is the 
SU(2)$_L$ metric, $h_1$ and $h_2$ are Yukawa coupling constants, and $\alpha$, 
$\beta$, and $\gamma$ are 
SU(3)$_c$ indices. Then there will be a one loop level contribution to 
a rank one finite, down quark mass 
matrix. In this loop the top quark mass will act as a seed and the loop 
will be completed with the propagators of the $\phi^{(1)}$ and $\phi^{(2)}$
fields mixed at tree-level by a term of the form
$\phi^{(1)}\phi^{(2)}\langle\phi_H\rangle\langle\phi_H\rangle$.

\subsection{Consistence with experimental constraints.}

In MI and MII flavor changing neutral currents (FCNC) 
are mediated by the new gauge boson 
$Z^\prime$\cite{koca,po} and by the Higgs fields $\phi_H,\phi^{(1)}$ and 
$\phi^{(2)}$. The FCNCs resulting from $Z^\prime$ can be suppressed by giving 
the new gauge boson a mass equal to or larger than 100 TeVs.
\vskip.3cm

The FCNCs resulting from Higgs couplings can also be suppressed by giving 
the Higgs bosons sufficiently large masses. Naively, a mass of 100 TeVs 
for each one of those scalar fields will be consistent with the experimental 
constraints. Since the scalar sector introduced up to this point serves 
only as a starting point for more realistic mass generation schemes,
we will not pursue any detailed Higgs phenomenology in 
the present paper.

\subsection{Masses for the other fermions and Mixing angles.}
Another concern of horizontal models of this kind is how to generate the 
radiative corrections that are assumed to provide the smaller masses and 
mixing angles in the models. For this purposes the 
{\it cascade} mechanism may be invoked. In this mechanism the light 
particles gain masses at various orders of perturbation theory from masses 
induced at the previous order of approximation. This mechanism requires the 
introduction of new scalars and is presented in detail in Refs.\cite{ma,he}. 
\vskip.3cm

An interesting feature of a particular version of the cascade
mechanism\cite{ma} is that it couples the up quark to the
strange and bottom quarks and couples the down quark to the
charm and top quarks through the higher order corrections, thus
providing a natural explanation for the observation m$_u<$m$_d$.
\vskip.3cm

\subsection{ Embedding in a higher symmetry model.}
In the two models under consideration the traces of the horizontal 
hypercharges  vanish
in the family
basis. Therefore U(1)$_H$ or U(1)$^{que} \otimes 
$U(1)$^{dl}$ can be embedded into a simple or semisimple group.
Rescaling the H hypercharges we can write  
the generators as

$$
T_3 = diag(1, 0, -1)
$$

\noindent
or as

$$
(1/2)\lambda_3 = diag(1/2, -1/2, 0)
$$

\noindent
corresponding to (one of) the diagonal generators of an SU(2)$_H$ 
or an SU(3)$_H$ group.
The higher symmetry horizontal group can then be SU(2)$_{H}$,  
SU(2)$^{que}\otimes$SU(2)$^{dl}$, SU(3)$_{H}$, or
SU(3)$^{que}\otimes$SU(3)$^{dl}$.
In any of these cases the model should contain additional features, such
as extra fermions, in order to avoid the constraints that lead us
to discard them.

\section{Conclusion}
The requirement of anomaly cancellation for the SM augmented with 
a horizontal factor,  and with no additional fermions, other than 
right handed neutrinos, constitutes an strong condition. Without forcing
by hand the orientation of the vacuum, the set of the viable G$_H$, groups
is limited to U(1)$_H$ and U(1)$_{H1}\otimes $U(1)$_{H2}$. In each case two 
models have been found  with the 
fermions having the horizontal hypercharges displayed in table 1b.

\section{Acknowledgements}
This work was partially supported by CONACyT in Mexico and
COLCIENCIAS in Colombia. One of us (A.Z) acknowledges the hospitality
of Prof. J. Bernabeu and 
of the Theory Group at the University of Valencia as well as the financial 
support 
of Direcci\'on General de Investigaci\'on Cient\'\i fica y T\'ecnica
(DGICYT) of the Ministry of Education and Science of Spain and the 
hospitality 
and financial support of the Institute of Nuclear Theory in Seattle during 
part of the summer. 
\pagebreak

\setcounter{equation}{0}
\renewcommand{\theequation}{A\arabic{equation}}

\noindent
{\Large\bf APPENDIX A. Irrational and complex solutions to the anomaly 
constraints}\\
\vskip.3cm

In this appendix we present a new set of solutions to the anomaly 
constraint equations for the gauge group 
SU(3)$_c\otimes$SU(2)$_L\otimes$U(1)$_Y\otimes$U(1)$_H$, for the
case when the anomalies are canceled by an interplay among
families. We look for solutions to Eqs. 1 to 6 where a 
sum over $i=1,2,3$ should be understood as indicated in section 2.2, 
and again we restrict to the case where  
$\sum_{i=1}^3{\rm Y}_{\eta i}=0$. We classify this new set of
solutions in the following classes:
\vskip.3cm

\noindent
\underline{CLASS A$^\prime$} With
Y$_{ui}$=Y$_{di}$=Y$_{qi}=0; i=1,2,3$ \\
The constraint equations are now
 
\[\sum_{i=1}^3 Y_{ei}= \sum_{i=1}^3 Y_{li}=0 \]

\[\sum_{i=1}^3 Y^2_{ei}=\sum_{i=1}^3 Y^2_{li}\equiv 2a\]

\begin{equation}
\sum_{i=1}^3 Y^3_{ei}=
-2\sum_{i=1}^3 Y^3_{li}\equiv -3b  \label{eq:primadas}
\end{equation}

\noindent
where $a$ and $b$ are arbitrary
numbers. This class and class A of section 2 overlap when b= 0, a = $\delta^2$, 
$\delta^{\prime}$ = 0.
\vskip.3cm

The roots of the cubic equation

\[ Y^3-aY + b=0\]

\noindent
satisfy eqs. \ref{eq:primadas} and their ratios are
 irrational or complex numbers for b $\neq 0$.
\vskip.3cm

Other classes are given by:
\vskip.3cm

\noindent
\underline{CLASS B$^\prime$} With
Y$_{ei}$=Y$_{ui}$=Y$_{di}$=0; i=1,2,3 \\
\vskip.3cm

\noindent
\underline{CLASS C$^\prime$} With
Y$_{qi}$=Y$_{ui}$=Y$_{ei}$=0; i=1,2,3, etc.
\vskip.3cm

The exotic solutions presented in this Appendix are not considered in the
main text because we do not envisage 
how to fit them  in a natural way into more general theories.

\pagebreak

\noindent
{\Large\bf APPENDIX B. Horizontal quantum numbers of Yukawa terms in the 
MI and MII Models.}
\vskip.3cm

I). MI model.
\vskip.3cm

$$
{\it \bf {\cal L}^{(2/3)}} \equiv \phi^{\dagger}u^c_{i}\left(
\begin{array}{c}u\\d\end{array}
\right)_j: \; \; \; \left(\begin{array}{ccc}
	   0             & -2\delta^{\prime}& -\delta^{\prime}\\
	-2\delta^{\prime}& -4\delta^{\prime}&-3\delta^{\prime}\\				
	-\delta^{\prime} & -3\delta^{\prime}&-2\delta^{\prime} \end{array}
        \right)
$$
\vskip.3cm

$$
{\it \bf {\cal L}^{(-1/3)}} \equiv \phi^{c\dagger}d^c_{i}\left(
\begin{array}{c}u\\d\end{array}
\right)_j: \; \; \; \left(\begin{array}{ccc}
\delta + 3\delta^{\prime} & \delta + \delta^{\prime}&\delta + 2\delta^{\prime}\\
-\delta + 3\delta^{\prime}&-\delta + \delta^{\prime}&-\delta+ 2\delta^{\prime}\\
         3\delta^{\prime} &          \delta^{\prime}&   2\delta^{\prime}\\
	\end{array}
        \right)
$$
\vskip.3cm

$$
{\it \bf {\cal L}^{(-1)}} \equiv \phi^{c\dagger}e^c_{i}\left(
\begin{array}{c}\nu \\e \end{array}
\right)_j: \; \; \; \left(\begin{array}{ccc}
\delta + 3\delta^{\prime} & -\delta + 3\delta^{\prime}& 3\delta^{\prime}\\
\delta +   \delta^{\prime}&-\delta + \delta^{\prime}  &  \delta^{\prime}\\
\delta + 2\delta^{\prime} &-\delta  + 2\delta^{\prime}& 2\delta^{\prime}\\
	\end{array}
        \right)
$$
\vskip.3cm

$$
{\it \bf {\cal L}^{(0)}} \equiv \phi^{\dagger}\nu ^c_{i}\left(
\begin{array}{c}\nu \\e \end{array}
\right)_j 
\left\{ 
\begin{array}{l} Y_{\nu i} = \delta_i :\; \; \; \left(\begin{array}{ccc}
   2\delta -2\delta^{\prime} & -2\delta^{\prime}&\delta -2\delta^{\prime}\\
    -2\delta^{\prime}&-2\delta-2\delta^{\prime}&-\delta-2\delta^{\prime}\\				
\delta-2\delta^{\prime} & -\delta-2\delta^{\prime}&-2\delta^{\prime} \end{array}
        \right)\\  \\
       Y_{\nu i} = 0:\; \; \; \left(\begin{array}{ccc}
  \delta -2\delta^{\prime}& -\delta-2\delta^{\prime}& -2\delta^{\prime}\\
  \delta -2\delta^{\prime}& -\delta-2\delta^{\prime}& -2\delta^{\prime}\\
  \delta -2\delta^{\prime}& -\delta-2\delta^{\prime}& -2\delta^{\prime}		
		\end{array}
        \right) \end{array} \right.
	$$

\vspace{1cm}

II). MII model.
\vspace{.3cm}

$$
{\it \bf {\cal L}^{(2/3)}} \equiv \phi^{\dagger}u^c_{i}\left(
\begin{array}{c}u\\d\end{array}
\right)_j: \; \; \; \left(\begin{array}{ccc}
	  4\delta^{\prime}& 2\delta^{\prime}& 3\delta^{\prime}\\
	2\delta^{\prime}&    0              &  \delta^{\prime}\\				
	3\delta^{\prime} & \delta^{\prime}& 2\delta^{\prime} \end{array}
        \right)
$$
\vspace{.3cm}

$$
{\it \bf {\cal L}^{(-1/3)}} \equiv \phi^{c\dagger}d^c_{i}\left(
\begin{array}{c}u\\d\end{array}
\right)_j: \; \; \; \left(\begin{array}{ccc}
\delta -\delta^{\prime} & \delta -3 \delta^{\prime}&\delta - 2\delta^{\prime}\\
-\delta-\delta^{\prime}& -\delta -3 \delta^{\prime}&-\delta- 2\delta^{\prime}\\
       -\delta^{\prime} &        -3 \delta^{\prime}&        -2\delta^{\prime}\\
	\end{array}
        \right)
$$
\vspace{.3cm}

$$
{\it \bf {\cal L}^{(-1)}} \equiv \phi^{c\dagger}e^c_{i}\left(
\begin{array}{c}\nu \\e \end{array}
\right)_j: \; \; \; \left(\begin{array}{ccc}
\delta -\delta^{\prime} & -\delta -\delta^{\prime}& -\delta^{\prime}\\
\delta -3 \delta^{\prime}&-\delta -3 \delta^{\prime}  & -3 \delta^{\prime}\\
\delta - 2\delta^{\prime} &-\delta  - 2\delta^{\prime}& -2\delta^{\prime}\\
	\end{array}
        \right)
$$
\vspace{.3cm}

$$
{\it \bf {\cal L}^{(0)}} \equiv \phi^{\dagger}\nu ^c_{i}\left(
\begin{array}{c}\nu \\e \end{array}
\right)_j \left\{ \begin{array}{l} Y_{\nu i} = \delta_i 
:\; \; \; \left(\begin{array}{ccc}
2\delta +2\delta^{\prime}&         2\delta^{\prime}&\delta +2\delta^{\prime}\\
         2\delta^{\prime}&-2\delta+2\delta^{\prime}&-\delta+2\delta^{\prime}\\				
\delta  +2\delta^{\prime}&- \delta+2\delta^{\prime}&        2\delta^{\prime} 
\end{array}
        \right)\\    \\ 
        		       Y_{\nu i} = 0:\; \; \; \left(\begin{array}{ccc}
  \delta +2\delta^{\prime}& -\delta+2\delta^{\prime}& 2\delta^{\prime}\\
  \delta +2\delta^{\prime}& -\delta+2\delta^{\prime}& 2\delta^{\prime}\\
  \delta +2\delta^{\prime}& -\delta+2\delta^{\prime}& 2\delta^{\prime}		
		\end{array}
        \right) 
\end{array} \right.
$$
\vskip1cm

Here, as well as in the main text, the contraction of spinor indices should
be understood taking into account the charge conjugation matrix C,

$$
u^cq \equiv u^{c T}_L C q_L, \; \; \; \; \; C = i\gamma_2 \gamma_0.
$$

\pagebreak

\pagebreak

\noindent
{Table 1a}. 
U(1)$_Y$, U(1)$_H$, U(1)$^L$ and U(1)$^R$ 
charges for the known fermions. $i=1,2,3$ is a flavor 
index denoting first, second and third family respectively. 
The Y$_{SM}$ values stated are family independent. All fermions are 
left handed.
\vskip.8cm

\begin{tabular}{||l|cccccc||}  \hline
 & $l_{i}=(\nu ,e)_{i}$ & $e^c_{i}$ & $q_{i}=$
(u,d)$_{i}$ & $u^c_{i}$ & $d^c_{i}$ & $\nu^c_{i}$ \\ \hline
Y$_{SM}$ & $-1$ & 2 & 1/3 & $-4/3$ & 2/3 & 0 \\
Y$_{H}$ & Y$_{li}$ & Y$_{ei}$ & Y$_{qi}$ & Y$_{ui}$ & 
Y$_{di}$ & Y$_{\nu i}$  \\ 
Y$^L$ & Y$_{li}^L$ & 0 & Y$_{qi}^L$ & 0 & 0 & 0  \\
Y$^R$ & 0 & Y$_{ei}^R$ & 0 & Y$_{ui}^R$ & Y$_{di}^R$ & 
Y$^R_{\nu i}$  \\  \hline
\end{tabular}
\vskip3cm

\noindent
{Table 1b}.
U(1)$_Y$ and U(1)$^{que}\otimes $U(1)$^{dl}$ [or U(1)$_H$ with Y$_H =
Y^{que} + Y^{dl}$]
charges, as 3 $\times$ 3 diagonal matrices in the family basis, for the 
known fermions in MI and MII models. 
The Y$_{SM}$ matrices are proportional to the 3 $\times$ 3 unit matrix.

\vspace{1cm}
\hspace{-2cm}
\begin{tabular}{||l|ccc|ccc||} \hline
& \multicolumn{3}{c|}{U(1)$^{que}$} & \multicolumn{3}{c||}{ U(1)$^{dl}$} \\ 
\cline{2-7}
& $q_{i}=(u, \, d)_{i}$ & $u^c_{i}$ &  $e^c_{i}$ & 
 $l_{i}=(\nu, \, e)_{i}$ & $d^c_{i}$ & $\nu^c_{i}$ \\ \hline
Y$_{SM}$ & $1/3$ & -4/3 & 2  & $-1 $ & 2/3 & 0 \\
Y$^{que}$ & ($\delta^\prime$, -$\delta^\prime$, 0) &
	    ($\delta^\prime$, -$\delta^\prime$, 0) &
	    ($\delta^\prime$, -$\delta^\prime$, 0) & 0 & 0 & 0 \\
Y$^{dl}$ & 0&0&0& ($\delta$, -$\delta$, 0) &
	    ($\delta$, -$\delta$, 0) & ($\delta$, -$\delta$, 0) or 0 \\
Y$_{H}$ & ($\delta^\prime$, -$\delta^\prime$, 0) &
	  ($\delta^\prime$, -$\delta^\prime$, 0) &
	    ($\delta^\prime$, -$\delta^\prime$, 0) &
	  ($\delta$, -$\delta$, 0) &
	  ($\delta$, -$\delta$, 0) &
	    ($\delta$, -$\delta$, 0) or 0 \\  \hline
\end{tabular}
\pagebreak

\noindent
{Table 2}.
Summary of tree-level mass terms for all the possible models
with SU(3)$_c\otimes$SU(2)$_L\otimes$U(1)$_Y\otimes$U(1)$_H$
symmetry group allowed by the indicated Higgs field 
hypercharge $Y_{\phi}$. 

\vspace{1cm}

\hspace{-3cm}
\begin{tabular}{||l|c|c|c||}  \hline 
Y$_{\phi}$ & CLASS A & CLASS B & CLASS C \\ \hline
2$\delta^\prime$ & (1,1)$_u$;(2,2)$_d$ & & (1,1)$_u$ \\
$-2\delta^\prime$ & $(2,2)_u;(1,1)_d$ & & (2,2)$_u$ \\
0 & $(1,2)_u;(2,1)_u(3,3)_u;(1,2)_d;(2,1)_d;(3,3)_d$ & $(3,3)_u;(3,3)_d$ & 
$(1,2)_u;(2,1)_u;(3,3)_u;(3,3)_d$ \\
$\delta^\prime$ & $(1,3)_u;(3,1)_u;(2,3)_d;(3,2)_d$ & 
$(3,1)_u;(3,2)_d$& $(1,3)_u;(3,1)_u;(2,3)_d$ \\
$-\delta^\prime$ & $(2,3)_u;(3,2)_u;(1,3)_d;(3,1)_d$ & 
$(3,2)_u;(3,1)_d$& $(2,3)_u;(3,2)_u;(1,3)_d$ \\
$\delta+\delta^\prime$ & & $(1,1)_u;(2,2)_d$ & $(2,2)_d$ \\
$-\delta+\delta^\prime$ & & $(2,1)_u;(1,2)_d$ & $(2,1)_d$ \\
$\delta-\delta^\prime$ & & $(1,2)_u;(2,1)_d$ & $(1,2)_d$ \\
$-\delta-\delta^\prime$ & & $(2,2)_u;(1,1)_d$ & $(1,1)_d$ \\
$\delta$ & & $(1,3)_u;(2,3)_d$ & $(3,1)_d$ \\
$-\delta$ & & $(2,3)_u;(1,3)_d$ & $(3,2)_d$ \\  \hline
\end{tabular}

\pagebreak

\noindent
{Table 3}.
Possible lepton SU(2)$_{HL}\otimes$SU(2)$_{HR}$ representation assignments, 
and their anomalies. [HL(R)]$^2$Y stands for [SU(2)$_{HL(R)}]^2$U(1)$_Y$. The 
representation for $l_{i}$ refers to the SU(2)$_{HL}$ group and the 
representation for $e^c_{i}$ refers to the SU(2)$_{HR}$ group.
\vskip1cm

\begin{tabular}{||l|cccccc||}  \hline
 & $l_{i}$ & $e^c_{i}$ & [HL]$^2$Y.
& [HR]$^2$Y. & 2$_L^{'s}$ & 2$_R^{'s}$ \\
 &  &  & anomaly & anomaly & &  \\ \hline
L$_1$ & 3 & 3 & $-8$ & 8 & 0 & 0 \\
L$_2$ & 3 & 2+1 & $-8$ & 2 & 0 & 1 \\
L$_3$ & 2+1 & 3 & $-2$ & 8 & 2 & 0 \\
L$_4$ & 2+1 & 2+1 & $-2$ & 2 & 2 & 1 \\
L$_5$ & 3 & 1+1+1 & $-8$ & 0 & 0 & 0 \\
L$_6$ & 1+1+1 & 3 & 0 & 8 & 0 & 0 \\
L$_7$ & 2+1 & 1+1+1 & $-2$ & 0 & 2 & 0 \\
L$_8$ & 1+1+1 & 2+1 & 0 & 2 & 0 & 1 \\
L$_9$ & 1+1+1 & 1+1+1 & 0 & 0 & 0 & 0 \\ \hline
\end{tabular}

\pagebreak

\noindent
{Table 4a}.
Possible quark SU(2)$_{HL}\otimes$SU(2)$_{HR}$ representation assignments, 
and their anomalies. [HL(R)]$^2$Y stands for [SU(2)$_{HL(R)}]^2$U(1)$_Y$. The 
representation for $q_{i}$ refers to the SU(2)$_{HL}$ group and the 
representation for $u^c_{i}$ and $d^c_{i}$ refers to the SU(2)$_{HR}$ 
group.
\vskip1cm

\begin{tabular}{||l|ccccccc||}  \hline
 & $q_{i}$ & $u^c_{i}$ & $d^c_{i}$ & 
[HL]$^2$Y & [HR]$^2$Y & 2$_L^{'s}$ & 2$_R^{'s}$ \\
 & & &  & anomaly & anomaly & &  \\ \hline
Q$_1$ & 3 & 3 & 3 & 8 & $-8$ & 0 & 0 \\
Q$_2$ & 3 & 2+1 & 3 & 8 & 4 & 0 & 3 \\
Q$_3$ & 3 & 3 & 2+1 & 8 & $-14$ & 0 & 3 \\
Q$_4$ & 2+1 & 3 & 3 & 2 & $-8$ & 6 & 0 \\
Q$_5$ & 3 & 2+1 & 2+1 & 8 & $-2$ & 0 & 6 \\
Q$_6$ & 2+1 & 2+1 & 3 & 2 & 4 & 6 & 3 \\
Q$_7$ & 2+1 & 3 & 2+1 & 2 & $-14$ & 6 & 3 \\
Q$_8$ & 2+1 & 2+1 & 2+1 & 2 & $-2$ & 6 & 6 \\
Q$_9$ & 3 & 1+1+1 & 3 & 8 & 8 & 0 & 0 \\
Q$_{10}$ & 3 & 3 & 1+1+1 & 8 & $-16$ & 0 & 0 \\
Q$_{11}$ & 1+1+1 & 3 & 3 & 0 & $-8$ & 0 & 0 \\
Q$_{12}$ & 3 & 1+1+1 & 1+1+1 & 8 & 0 & 0 & 0 \\
Q$_{13}$ & 1+1+1 & 1+1+1 & 3 & 0 & 8 & 0 & 0 \\
Q$_{14}$ & 1+1+1 & 3 & 1+1+1 & 0 & $-16$ & 0 & 0 \\
Q$_{15}$ & 2+1 & 1+1+1 & 2+1 & 2 & 2 & 6 & 3 \\
Q$_{16}$ & 2+1 & 2+1 & 1+1+1 & 2 & $-4$ & 6 & 3 \\
Q$_{17}$ & 1+1+1 & 2+1 & 2+1 & 0 & $-2$ & 0 & 6 \\
Q$_{18}$ & 2+1 & 1+1+1 & 1+1+1 & 2 & 0 & 6 & 0 \\
Q$_{19}$ & 1+1+1 & 1+1+1 & 2+1 & 0 & 2 & 0 & 3 \\
Q$_{20}$ & 1+1+1 & 2+1 & 1+1+1 & 0 & $-4$ & 0 & 3 \\
Q$_{21}$ & 1+1+1 & 1+1+1 & 1+1+1 & 0 & 0 & 0 & 0 \\
Q$_{22}$ & 3 & 2+1 & 1+1+1 & 8 & $-4$ & 0 & 3 \\
Q$_{23}$ & 3 & 1+1+1 & 2+1 & 8 & 2 & 0 & 3 \\
Q$_{24}$ & 2+1 & 3 & 1+1+1 & 2 & $-16$ & 6 & 0 \\
Q$_{25}$ & 2+1 & 1+1+1 & 3 & 2 & 8 & 6 & 0 \\
Q$_{26}$ & 1+1+1 & 3 & 2+1 & 0 & $-14$ & 0 & 3 \\
Q$_{27}$ & 1+1+1 & 2+1 & 3 & 0 & 4 & 0 & 3 \\  \hline
\end{tabular}

\pagebreak

\noindent
Table 4b. Rank of the up-quark mass matrix for the different representation 
contents of $q$, $u$ and $\phi$ under some SU(2)$_{H1}$ group. Q$_i$ is one
of the quark arrangements with the same $n_q$ and $n_u$ content which helps 
in finding the corresponding case in tables IV and V of Ref. \cite{volkas}.

\vskip1cm

\hspace{2cm}
\begin{tabular}{||cclcc||}  \hline
$n_q$   & $n_u$   & Q$_i$    & $n_{\phi}$ & rank \\ \hline \hline
1+1+1 & 1+1+1 & Q$_{21}$ &  1       &  3    \\ \hline
1+1+1 & 2+1   & Q$_{17}$ &  1       &  1    \\ 
      &       &          &  2       &  1   \\ \hline
1+1+1 &  3    & Q$_{11}$ &  3       &  1    \\ \hline
2+1   & 1+1+1 & Q$_{18}$ &  1       &  1    \\ 
      &       &        &  2       &  1    \\ \hline
2+1   & 2+1   & Q$_{6}$  &  1       &  3    \\ 
      &       &        &  2       &  2     \\
      &       &        &  3       &  2    \\ \hline
2+1   & 3     & Q$_{4}$  &  2       &  2    \\ 
      &       &        &  3       &  1    \\
      &       &        &  4       &  2    \\ \hline
3     & 1+1+1 & Q$_{12}$ &  3       &  1    \\ \hline
3     & 2+1   & Q$_{5}$  &  2       &  2    \\ 
      &       &        &  3       &  1    \\
      &       &        &  4       &  2    \\ \hline
3     & 3     & Q$_{10}$ &  1       &  3    \\ 
      &       &        &  3       &  2    \\
      &       &        &  5       &  3    \\ \hline
\end{tabular}

\vskip1cm

\noindent
Table 4c. Quark and Higgs SU(2)$_{H1}$ representations with a rank-one
up-quark mass matrix and with simultaneous
rank-zero down-quark mass matrix and the corresponding 
[SU(2)$_{H1}$]$^2$ U(1)$_Y$ anomaly.
\vskip.3cm

\hspace{1.5cm}
\begin{tabular}{||ccccc||} \hline
$n_q$    & $n_u$   & $n_d$  & $n_{\phi}$ & [SU(2)$_{H1}$]$^2$ U(1)$_Y$ \\ 
\hline \hline
1+1+1  & 2+1   & 3    &  1       &  4  \\ \hline
1+1+1  & 2+1   & 1+1+1&  2       & -4  \\
1+1+1  & 2+1   &  3   &  2       &  4  \\ \hline
1+1+1  & 3     & 1+1+1&  3       &  -16  \\ 
1+1+1  & 3     & 2+1  &  3       &  -14  \\ \hline
2+1    & 1+1+1 &  3   &  1       &  10  \\ \hline
2+1    & 3     & 1+1+1&  3       &  -14  \\ \hline
\end{tabular}

\pagebreak

\noindent
{Table 5}.
Possible quark and lepton SU(3)$_H$ representations which are free of the 
gauge SU(3)$_H$ anomaly, where a 1 value stands for $1+1+1$. 
[3H]$^2$Y stands for the [SU(3)$_H]^2$U(1)$_Y$ anomaly 
value. The number of representations can be doubled by the replacement 
$3\leftrightarrow\bar{3}$, but the new 18 arrangements are equivalent to the 
present ones (as far as the low energy phenomenology is concerned). 
\vskip1cm

\hspace{2cm}
\begin{tabular}{||l|ccccccc||}  \hline
 & $e^c_{i}$ & $\nu^c_{i}$ & $l_{i}$
 & $u^c_{i}$ & $d^c_{i}$  & $q_{i}$ 
 & [3H]$^2$Y \\ \hline
M$_1$ & 3 & $\bar{3}$ & 1 & 1 & 1 & 1 & 2 \\
M$_2$ & 1 & 1 & 1 & 3 & $\bar{3}$ & 1 & $-2$\\
M$_3$ & 3 & 3 & $\bar{3}$ & 1 & 1 & 1 & 0 \\
M$_4$ & 1 & 1 & 1 & 3 & 3 & $\bar{3}$ & 0 \\
M$_5$ & 3 & 1 & 3 & $\bar{3}$ & 1 & 1& $-4$ \\
M$_6$ & 3 & 1 & 3 & 1 & $\bar{3}$ & 1 & 2 \\
M$_7$ & 1 & 3 & 3 & $\bar{3}$ & 1 & 1 & $-6$ \\
M$_8$ & 1 & 3 & 3 & 1 & $\bar{3}$ & 1 & 0 \\
M$_9$ & 3 & 1 & 3 & 3 & 1 &  $\bar{3}$ & $-2$ \\
M$_{10}$ & 3 & 1 & 3 & 1 & 3 & $\bar{3}$ & 4 \\
M$_{11}$ & 1 & 3  & 3 & 3 & 1 & $\bar{3}$ & $-4$ \\
M$_{12}$ & 1 & 3 & 3 &1 & 3 & $\bar{3}$ & 2 \\
M$_{13}$ & 3 & $\bar{3}$ & 1 & $\bar{3}$ & 3 & 1 & 0 \\
M$_{14}$ & $\bar{3}$ & 3 & 1 & $\bar{3}$ & 3 & 1 & 0 \\
M$_{15}$ & $\bar{3}$ & 3 & 1 & 3 & 3 & $\bar{3}$ & 2 \\
M$_{16}$ & 3 & $\bar{3}$ & 1 & 3 & 3 & $\bar{3}$ & 2 \\
M$_{17}$ & 3 & 3 & $\bar{3}$ & 3 & 3 & $\bar{3}$ & 0 \\
M$_{18}$ & $\bar{3}$ & $\bar{3}$ & 3 & 3 & 3 & $\bar{3}$ & 0 \\  
M$_{19}$ & 1 & 1 & 1 & 1 & 1 & 1 & 0 \\ \hline
\end{tabular}


\begin{thebibliography}{99}

\bibitem{hsh}
R.Barbieri and D.V.Nanopoulos, Phys. Lett. B {91} (1980) 369 ; {95B}, 
(1980) 43.


\bibitem{mhsh}
W.A.Ponce, A.Zepeda, A.H.Galeana and R.Mart\'\i nez, Phys. Rev {D44}
(1991) 2166.


\bibitem{po}
W.A.Ponce, Phys. Rev. {D36} (1987) 962; 

W.A.Ponce, A. Zepeda, and J.M. Mira, Z.Phys. C, Vol. 69, No 4, February 1996,
in press. 


\bibitem{anom}
S.L.Adler, Phys. Rev. {177} (1969) 2426; 69, No4, February 1996

J.S.Bell and R.Jackiw, Nuovo Cimento {51} (1969) 47; 

See also ``{\it Current Algebra and Anomalies}"
by S.B. Treiman, R Jackiw, B. Zumino, and E. Witten. World Scientific, Singapore,
1985.


\bibitem{del}
A.Salam and R.Delburgo, Phys. Lett. {40B} (1972) 381; 

L.Alvarez Gaum\'e and E.Witten, Nucl. Phys. {B234} (1983) 269.


\bibitem{framoh} P.H. Frampton and R.N. Mohapatra, Phys. Rev. D 50 (1994) 3569.

\bibitem{papa}
M.Green and J.Schwarz, Phys. Lett. {B149} (1984) 117.

\bibitem{ibanezross}
L.Ib\'a\~nez and G.G. Ross, Nucl. Phys. B 332 (1994) 100;

E. Papageorgiu, Z. Physik C 64 (1994) 509;

P. Binetruy and  P. Ramond, Phys. Lett. B 350 (1995) 49;

V. Janin and R. Shrock, ``{\it Models of Fermion Mass Matrices based on 
Flavor and Generation-Dependent U(1) Gauge Symmetry}", 1994;

Y. Nir, ``{\it Gauge Unification, Yukawa Hierarchy and the $\mu$ Problem}",
hep-ph 9504312.


\bibitem{koca}
A.Davidson, M.Koca and K.C.Wali, Phys. Rev. Lett. {43} (1979) 92; 
Phys. Rev. {D20} (1979) 1195; {D21} (1980) 787.


\bibitem{seesaw}
M.Gell-Mann, P. Ramond, and R. Slansky, in ``{\it Supergravity}", proceedings 
of the workshop, Stony Brook, New York 1979, edited by P. van Nieuwenhuizen 
and D.Z. Freedman (North-Holland, Amsterdam 1979), p. 315; 

T. Yanahida, in 
``{\it Proceedings of the workshop on Unified Theories and the Baryon Number 
in the Universe}", edited by A. Sawada and A. Sugamoto (KEK report No. 79-18, 
Tsukub-Gun, Ibaraki-Ken, Japan, 1979).

\bibitem{volkas}
D.S.Shaw and R.R.Volkas, Phys. Rev. {D47} (1993) 241.

\bibitem{witten}
E.Witten, Phys. Lett. {B117}(1982) 324.

\bibitem{yana}
T.Yanahida, Phys. Rev. {D20} (1979) 2986.

\bibitem{zupa}
G.Zoupanos, Z.Phys. {C11} (1981) 27; 

Phys. Lett. {115B} (1982) 221;

E.Papantanopoulos and G.Zoupanos, Phys. Lett. {110B} (1982) 465; 
Z.Phys. {C16} (1983) 361.

\bibitem{ray}
K.Bandyopadhyay and A.K.Ray, Phys. Rev. {D38} (1988) 2231.

\bibitem{bere}
Z.G.Berezhiani and M.Y. Khlopov, Z.Phys. {C49} (1991) 73.

\bibitem{ma}
E.Ma, Phys. Rev. Lett {64} (1990) 2866.

\bibitem{he}
B.S.Balakrishna, Phys. Rev. Lett {60} (1988) 1602; 

X.G.He, R.R.Volkas and D.D.Wu, Phys. Rev. {D41} (1990) 1630.

\end{thebibliography}
\end{document}